\documentclass[man, floatsintext, a4paper]{apa6}

\usepackage[american]{babel}
\usepackage[utf8]{inputenc}
\usepackage{graphicx}
\usepackage{csquotes}
\usepackage{amsmath}
\usepackage{amsthm}
\usepackage{amssymb}
\usepackage{bm,array}
\usepackage{latexsym}

\usepackage{microtype}

\usepackage{color}
\usepackage[]{hyperref}

\usepackage{natbib}
\theoremstyle{definition}
\newtheorem{defn}{Definition}

\title{Model selection by minimum description length: Lower-bound sample sizes for the Fisher information approximation}

\shorttitle{Fisher information approximation}

\author{Daniel W. Heck, Morten Moshagen, and Edgar Erdfelder}
\affiliation{University of Mannheim}

\threeauthors{Daniel W. Heck}{Morten Moshagen}{Edgar Erdfelder}
\threeaffiliations{University of Mannheim}{University of Kassel}{University of Mannheim\\
\footnotesize
\flushleft\color{blue} \vspace{1cm}
For the published version of the article, see:
Heck, D. W., Moshagen, M., \& Erdfelder, E. (2014). 
Model selection by minimum description length: 
Lower-bound sample sizes for the Fisher information approximation. 
\textit{Journal of Mathematical Psychology, 60}, 29-34. 
\href{https://doi.org/10.1016/j.jmp.2014.06.002}{doi:10.1016/j.jmp.2014.06.002}
\vspace{-1cm}}

\leftheader{Heck, Moshagen, \& Erdfelder}

\authornote{
Daniel W. Heck, School of Social Sciences, Department of Psychology, University of Mannheim, Mannheim, Germany; Morten Moshagen, Department of Psychology, University of Kassel, Kassel, Germany; Edgar Erdfelder, Department of Psychology, School of Social Sciences, University of Mannheim, Mannheim, Germany.

Correspondence concerning this article should be addressed to Daniel W. Heck, Department of Psychology, School of Social Sciences, University of Mannheim, Schloss EO 254, D-68131 Mannheim,  Germany. E-mail: dheck@mail.uni-mannheim.de; Phone: (+49) 621 181 2145.

}

\abstract{The Fisher information approximation (FIA) is an implementation of the minimum description length principle for model selection. Unlike information criteria such as AIC or BIC, it has the advantage of taking the functional form of a model into account. Unfortunately, FIA can be misleading in finite samples, resulting in an inversion of the correct rank order of complexity terms for competing models in the worst case. As a remedy, we propose a lower-bound $N'$ for the sample size that suffices to preclude such errors. We illustrate the approach using three examples from the family of multinomial processing tree models.}

\keywords{Fisher information approximation; minimum description length; normalized maximum likelihood; model selection.}

\begin{document}

\maketitle

\section{1. Model selection and minimum description length}
In selecting a model from a set of competing models, a trade-off between the fit of a model and its complexity has to be made. On the one hand, a good model should describe observed data reasonably well; on the other hand, it should be as simple as possible so that results generalize beyond the current set of data. Flexible models with many parameters tend to fit too much noise beyond systematic patterns and hence might not predict new data well, a phenomenon known as overfitting \citep{Myung2000}. Consequently, reflecting the principle of Occam’s razor, if a less flexible model can account for the data equally well as a more complex model, the simple model is preferred, thus ensuring a high level of generalizability.

Implementations of Occam’s razor include well-known information criteria such as the Akaike information criterion \citep[AIC;][]{Akaike1973} or the Bayesian information criterion \citep[BIC;][]{Schwarz1978}. Both indices weigh the fit of a model in terms of the maximized likelihood against its complexity as measured by the number of free parameters. These criteria share the drawback that merely counting the number of free parameters fails to address the functional form of a model appropriately, that is, the structure of how the parameters are connected to each other. For example, in the context of psychophysics, \citet{Pitt2002} showed that Steven’s power law is more complex than Fechner’s logarithmic law even though both comprise the same number of free parameters. Likewise, taking the functional complexity into account is of fundamental importance when comparing models involving order constraints. Obviously, a model with order restrictions on the parameter vector $\boldsymbol\theta$ is less complex than a corresponding unrestricted model. However, neither AIC nor BIC can account for this difference because order restrictions do not affect the number of parameters of a model.

To overcome this limitation, \citet{Grunwald2000} proposed to rely on the principle of minimum description length (MDL) when selecting among competing models. This approach was developed in the field of algorithmic coding theory \citep{Rissanen1978} and addresses the issue of how much a given set of data can be compressed by a model. A model is preferred if it covers regularities in the data by means of the shortest code length \citep{Grunwald2007}. In the extreme case of randomly generated data, no compression is possible at all. In contrast, if data are generated deterministically without any noise, the code to describe these data can be shortened dramatically by giving the rule which generated the data \citep{Grunwald2005}. Models that compress data tightly provide a high level of generalizability, because they cover systematic patterns of the data that are likely to occur in future data as well. An implementation of the MDL principle was provided by \citet{Rissanen2001} who derived the normalized maximum likelihood (NML) to measure the stochastic complexity of a model given a data set,
\begin{equation}
\text{NML}=-\text{LML} + C_{\text{NML}}(N),
\end{equation}
where the maximum log-likelihood (LML) as a measure of fit is weighted against the complexity term
\begin{equation}
C_{\text{NML}}(N) = \ln \int_{\mathcal{X}^N} f(\mathbf{x} | \boldsymbol{\hat{\theta}}(\mathbf{x}))d\mathbf{x}.
\label{cnml}
\end{equation}
This complexity term is the natural logarithm of the integral over the maximum likelihoods across the whole outcome space $\mathcal X^N$ of potentially observable vectors $\mathbf x$ with number of observations $N$. Accordingly, a complex model that fits a wide range of observable data vectors will have a large value of $C_{\text{NML}}(N)$ compared to a model that fits only a small subset of observable data \citep{Myung2006}. Unfortunately, there is no general closed-form expression of $C_{\text{NML}}(N)$ and numerical estimation techniques such as Monte Carlo (MC) integration are often too time intensive for practical purposes. An alternative is the Fisher information approximation \citep[FIA;][]{Rissanen1996},
\begin{equation}
\text{FIA}=-\text{LML} + C_{\text{FIA}}(N),
\label{fia}
\end{equation}
which is asymptotically equivalent to NML. The complexity term $C_{\text{FIA}}(N)$ covers the number of free parameters $S$ and the number of observations $N$ in the first summand and considers the functional form of the model in the second,
\begin{equation}
C_{\text{FIA}}(N) = \frac S 2 \ln \left(\frac{N}{2 \pi}\right) + \ln \int_{\boldsymbol\Omega} \sqrt{ |\mathbf I(\boldsymbol \theta) | } d \boldsymbol \theta,
\label{cfia}
\end{equation}
where $\mathbf I(\boldsymbol \theta)$ is the Fisher information matrix of sample size one. This matrix contains the expected values of the second partial derivatives of the likelihood function and thereby captures the functional form of the model. Since the Fisher information matrix is often available in closed form, the integral in the second part of $C_{\text{FIA}}(N)$ is more tractable than the integral in $C_{\text{NML}}(N)$. $C_{\text{FIA}}(N)$ can be estimated by means of MC integration \citep{Pitt2002}.

\section{2. FIA in finite samples}
Although FIA approaches NML asymptotically, both measures can deviate substantially in finite samples. For hierarchical model families, in particular, \citet{Navarro2004} noted that the FIA complexity term for a nested model can become larger than that of a nesting model. Such a rank order reversal is obviously problematic, as a nested model, by definition, must be associated with a smaller complexity value. This notion is adequately reflected by NML, because the integral of the maximum likelihoods across the data space is always smaller for the nested model. Using FIA, however, such inverted rank orders of the complexity terms may occur, in turn resulting in a biased model selection in favor of the nesting model, as it always fits at least as well as the nested model. \citet{Navarro2004} supposed that the source of this problem is related to possible violations of the requirement that the maximum likelihood estimator $\boldsymbol{\hat\theta}$ must lie sufficiently within the model manifold, an assumption underlying the approximation of NML by FIA. However, the precise conditions under which FIA results are misleading are not yet established. 

Inversions of the rank order of FIA complexity terms for small $N$ may not only pose problems for model selection in nested model families but also for model selection in the more general class of NML stable models.
\begin{defn}[NML stability]
A set of stochastic models is called NML stable if the rank order of NML complexities $C_\text{NML}(N)$ across these models is identical for all $N \in \mathbb N$.
\label{nml-stable}
\end{defn}
The definition of $C_\text{NML}(N)$ directly implies NML stability for all pairs of nested models, as the NML complexity of a nested model is smaller than that of the nesting model for all sample sizes. However, NML stability does not necessarily hold for sets of non-nested models in general. As an example for a rank order inversion of NML complexities in non-nested model families, consider a hierarchical model that assigns a separate set of parameters to each participant. If such a model is compared to a different (non-nested) model that assumes a constant set of parameters for all $N$, the NML complexity of the latter model compared to the hierarchical model might indeed be larger for small samples, but smaller for larger samples.\footnote{Thanks to Dan Navarro for this example.} However, considering typical sets of non-nested models each having a constant number of parameters and presupposing independent and identically distributed data, NML stability is a reasonable assumption. In Section 4, we provide an example of an NML stable, non-nested model family with a FIA complexity rank order that deviates from the NML rank order for small $N$’s. Note that ---  unlike for nested models --- an inverted FIA rank order for non-nested models does not necessarily imply selection of the model with a larger NML complexity, because an overly small $C_\text{FIA}(N)$ can be compensated by a larger negative log-likelihood (cf. Eq. \ref{fia}). Nevertheless, in such a setting, FIA-based model selection will also be biased towards the more complex model.

To avoid biases in model selection using FIA for NML stable models, we propose to check whether the $C_\text{FIA}(N)$ rank order of the candidate models is invariant across different numbers $N$ of observations. Based on the definition of FIA in  (\ref{cfia}) it is easy to show that for any two models with a fixed but unequal number of parameters $S_i$ and $S_j$, respectively, the $C_\text{FIA}(N)$ rank order cannot be identical for all possible sample sizes. Since the integral in (\ref{cfia}) is independent of $N$, it is straightforward to determine the (single) sample size $N'_{i,j}$ for which the complexity terms of two models with $S_i \neq S_j$ are equal. Equating the FIA terms of two models $i$ and $j$ and solving for $N$ yields
\begin{equation}
N'_{i,j} = 2 \pi \exp \left[ \frac {2}{S_i - S_j} \left( \ln\int_{\boldsymbol{\Omega_j}} \sqrt{|\mathbf{I_j}(\boldsymbol\theta)|}d\boldsymbol\theta-\ln \int_{\boldsymbol{\Omega_i}}\sqrt{|\mathbf{I_i}(\boldsymbol\theta) |}d \boldsymbol\theta\right) \right].
\label{nij}
\end{equation}
When $N  > N'_{i,j}$, the $C_{\text{FIA}}(N)$ terms of the two competing models $i$ and $j$ will always result in the same rank order. Because $C_{\text{FIA}}(N)$ approximates $C_{\text{NML}}(N)$ for increasing $N$, this must be the correct (i.e., NML-consistent) rank order. By implication, for any $N < N'_{i,j}$ the rank order of complexity terms is incorrectly inverted. 

\begin{figure}[!ht]
   \includegraphics[keepaspectratio,width=16cm]{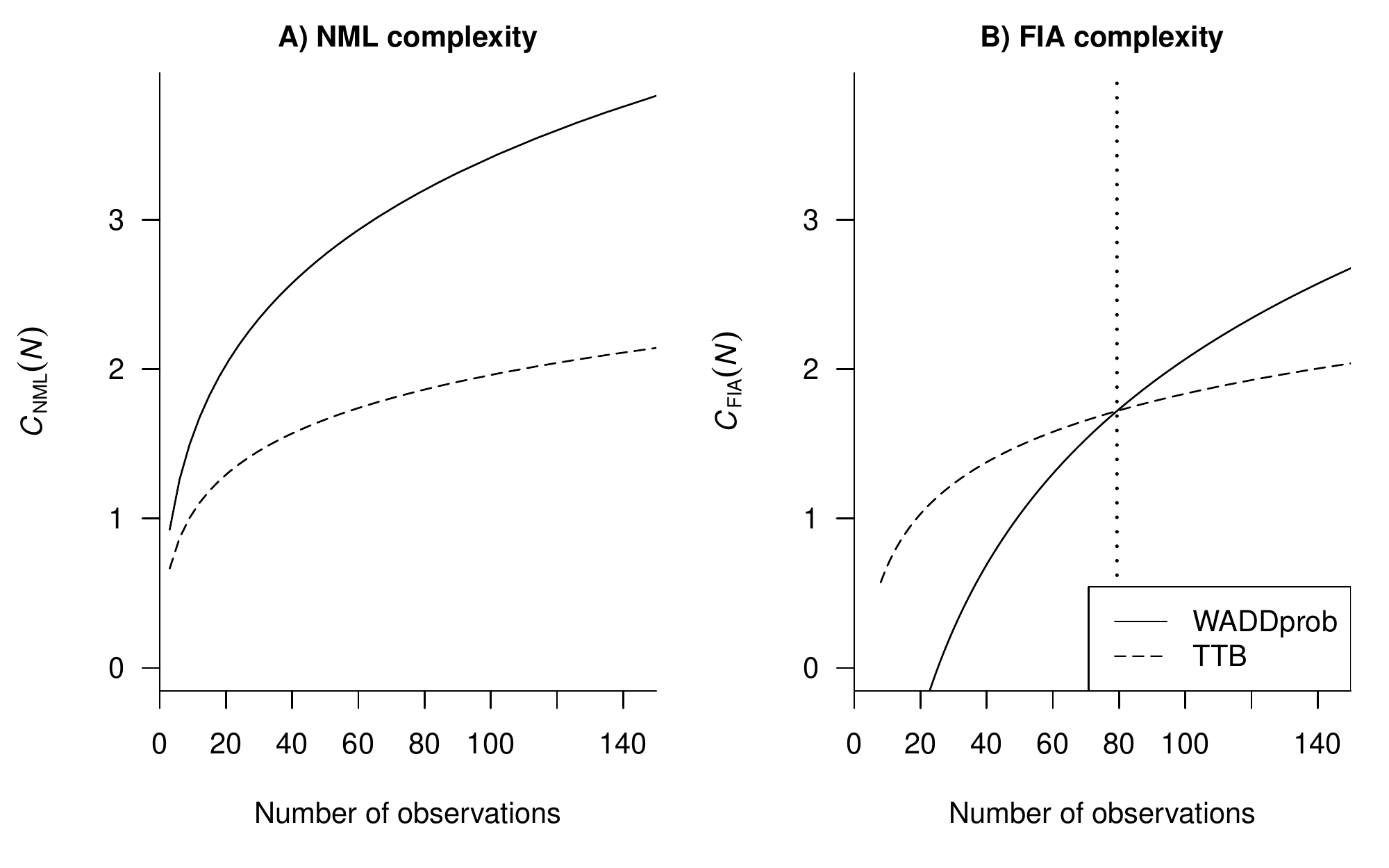}
   \caption{NML and FIA complexities for two decision strategies, take-the-best (TTB) and a probabilistic weighted-additive rule (WADDprob; Figure \ref{Fjdm}), as functions of the number of observations $N$. The dotted vertical line for FIA marks the lower bound \protect{$N' = 80$}.}
   \label{Fcfia}
\end{figure}

When plotting the FIA complexity terms for two models as logarithmic functions of $N$, $N'_{i,j}$ gives the number of observations at which these two lines intersect. Figure \ref{Fcfia}B illustrates this for two exemplary models that will be discussed in more detail in Section 4 of this article. Ideally, $N'_{i,j}$ should be as small as  possible thus ensuring the absence of the aforementioned bias for a large range of sample sizes. For two models with the same number of free parameters $S_i = S_j$, the complexity curves never intersect since the FIA complexity penalties can only differ in their intercepts --- the lines are either parallel or identical. In such a case, $N'_{i,j}$ is undefined and reversals of $C_{\text{FIA}}$ rank orders cannot occur.

Consequently, to make sure that the correct rank order of FIA complexity terms is found for a candidate set of two or more NML stable models, the actual number of observations in a study should exceed $N'_{i,j}$ for all pairs of models $(i, j)$ in the competition:

\begin{defn}[Lower-bound $N'$]
Considering a set of $M \in \mathbb N$ NML stable models, the lower-bound sample size $N'$ for the application of FIA to this set is 
\begin{equation}
N':=\max_{i,j \in \{1,...,M\}} N'_{i,j}, 
\end{equation}
with $N'_{i,j}$ given by (\ref{nij}) if $S_i \neq S_j$ and $N'_{i,j} = 0$ otherwise.
\end{defn}

\section{3. Nested model comparison}
We demonstrate the relevance of our approach using examples from the family of multinomial processing tree (MPT) models \citep{Batchelder1999, Erdfelder2009}. The models differ in size, thus showing that inverted rank orders in $C_{\text{FIA}}(N)$ can occur in various situations. An algorithm for the computation of the FIA complexity term for MPT models by means of MC integration was proposed by \citeauthor*{Wu2010} \citep[2010; for implementations see also][]{Moshagen2010, Singmann2013}. In all examples, we estimated the integral of the FIA complexity term based on one million samples.

\begin{figure}[!ht]
   \includegraphics[keepaspectratio,width=14cm]{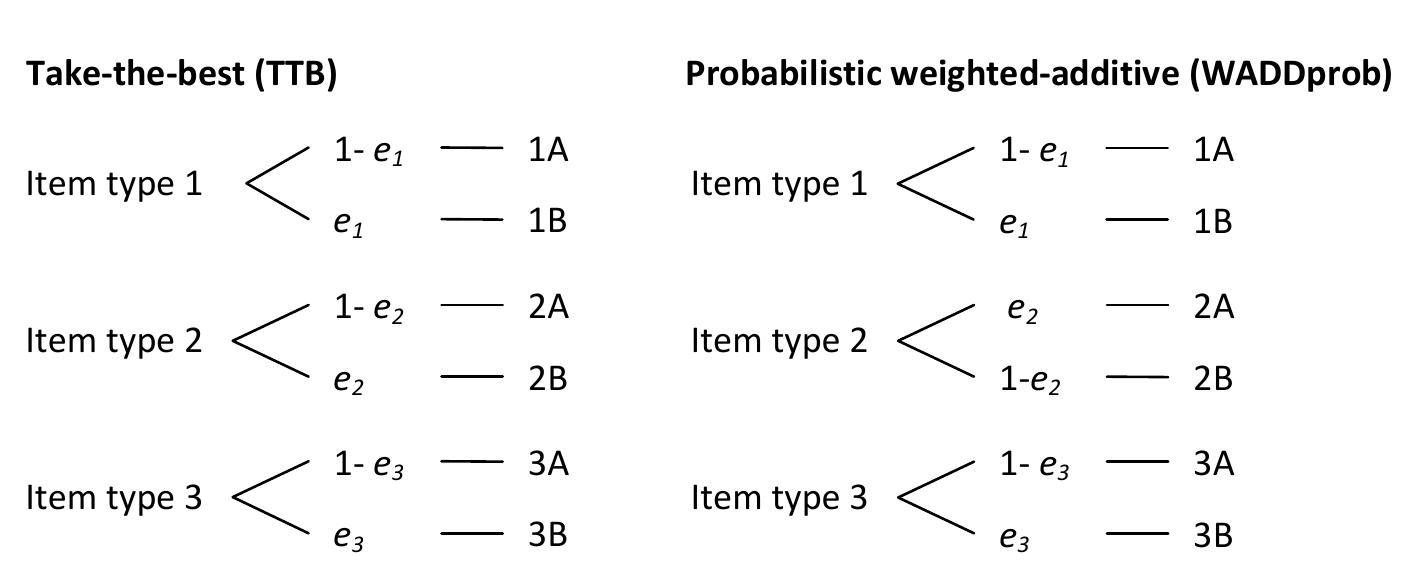}
   \caption{Decision strategies as used in \protect\citet{HilbigUR}.}
   \label{Fjdm}
\end{figure}

\begin{figure}[!ht]
   \includegraphics[keepaspectratio,width=14cm]{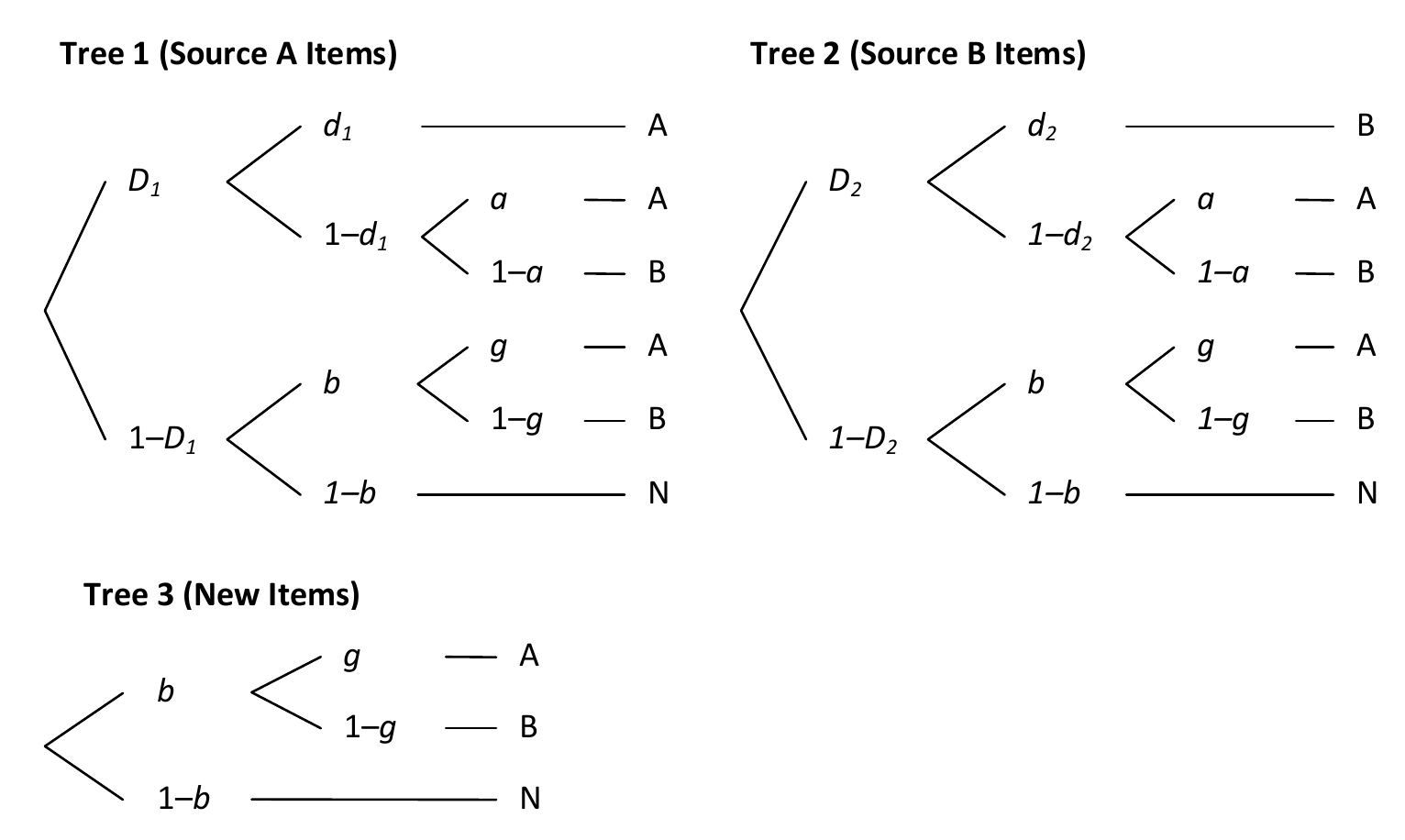}
   \caption{Source-monitoring model by \protect\citet{Batchelder1990}.}
   \label{Fsource}
\end{figure}

The first model is the one-high-threshold model of source monitoring as introduced by \citet[Figure \ref{Fsource}]{Batchelder1990} to disentangle the effects of item recognition ($D_1$ and $D_2$) and source discrimination ($d_1$ and $d_2$). In the nesting model, the item detection parameters ($D_1 = D_2$) and two of the guessing parameters ($a = g$) are set equal (model 5b). One nested model (model 4) additionally assumes equal source discrimination for both sources ($d_1 = d_2$). For this model pair, \citet{Wu2010} observed an inversion of the rank order of the FIA complexity penalties for an extreme proportion of 5\% new items and $N$ = 1,000. Using the proposed lower-bound $N'$, we can predict that this inversion vanishes when $N$ exceeds $N' = 1,393$. Importantly, in case of MPT models, both $C_\text{NML}(N)$ and $C_\text{FIA}(N)$ can vary depending on the proportion of observations per tree even when the overall $N$ remains constant. The effect of the proportion of new items on $N’$ is shown in Table \ref{tabN}. It is evident that a minimum is reached for a proportion of 50\% new items ($N'=292$), while choosing more extreme proportions leads to an increase in $N’$. Note that this minimum $N'$ is still larger than the number of observations for thought-disordered and non-thought-disordered manic participants ($n = 240$ each) in a data set by Harvey (1985) as discussed in \citet{Batchelder1990}. By implication, if these data were reanalyzed by means of FIA, the more complex nesting model 5b would always be preferred over the less complex nested model 4, irrespectively of the data. Clearly, FIA-based model selection would be severely misleading in such a case.

\newcolumntype{C}{>{\centering\arraybackslash}p{5em}}
\newcolumntype{x}[1]{%
{\raggedleft\hspace{0pt}}p{#1}}
\begin{table}
  \begin{threeparttable}  
  \caption{Lower-bound $N'$ for three model families as a function of the percentage of new items (in source monitoring models), distractor statements (in the ‘Who-said-what?’ paradigm), and Type-3 items (in the decision strategies model).}
  \label{tabN}
    \begin{tabular}{lrrrrr}         
    \midrule
    Model family   & \multicolumn{5}{c}{Proportion of new items, distractor} \\
    				   & \multicolumn{5}{c}{statements, and items type 3, respectively} \\  
    				   \cmidrule(r){2-6}
                            & \multicolumn{1}{c}{10\%} & \multicolumn{1}{c}{30\%}  
                            & \multicolumn{1}{c}{50\%} & \multicolumn{1}{c}{70\%}   
                            & \multicolumn{1}{c}{90\%}   \\ 
    \midrule
     Source-monitoring  &   \hspace{0.6cm}  750  &  \hspace{0.6cm} 340 & \hspace{0.6cm}  292&   \hspace{0.6cm} 344 &  \hspace{0.6cm} 770    \\
        'Who said what?'  &  8,616  & 3,415 & 2,568&  2,711 & 5,114\\
        Decision strategies & 108  &   80 &   87&   122 &  323   \\ 
    \midrule
    \end{tabular}
    \begin{tablenotes}[para,flushleft]
        {\small
            \textit{Note.} The ratio of the observations for the remaining item types is held constant at 1:1.

         }
    \end{tablenotes}
  \end{threeparttable}
\end{table}

\begin{figure}[!ht]
   \includegraphics[keepaspectratio,width=14cm]{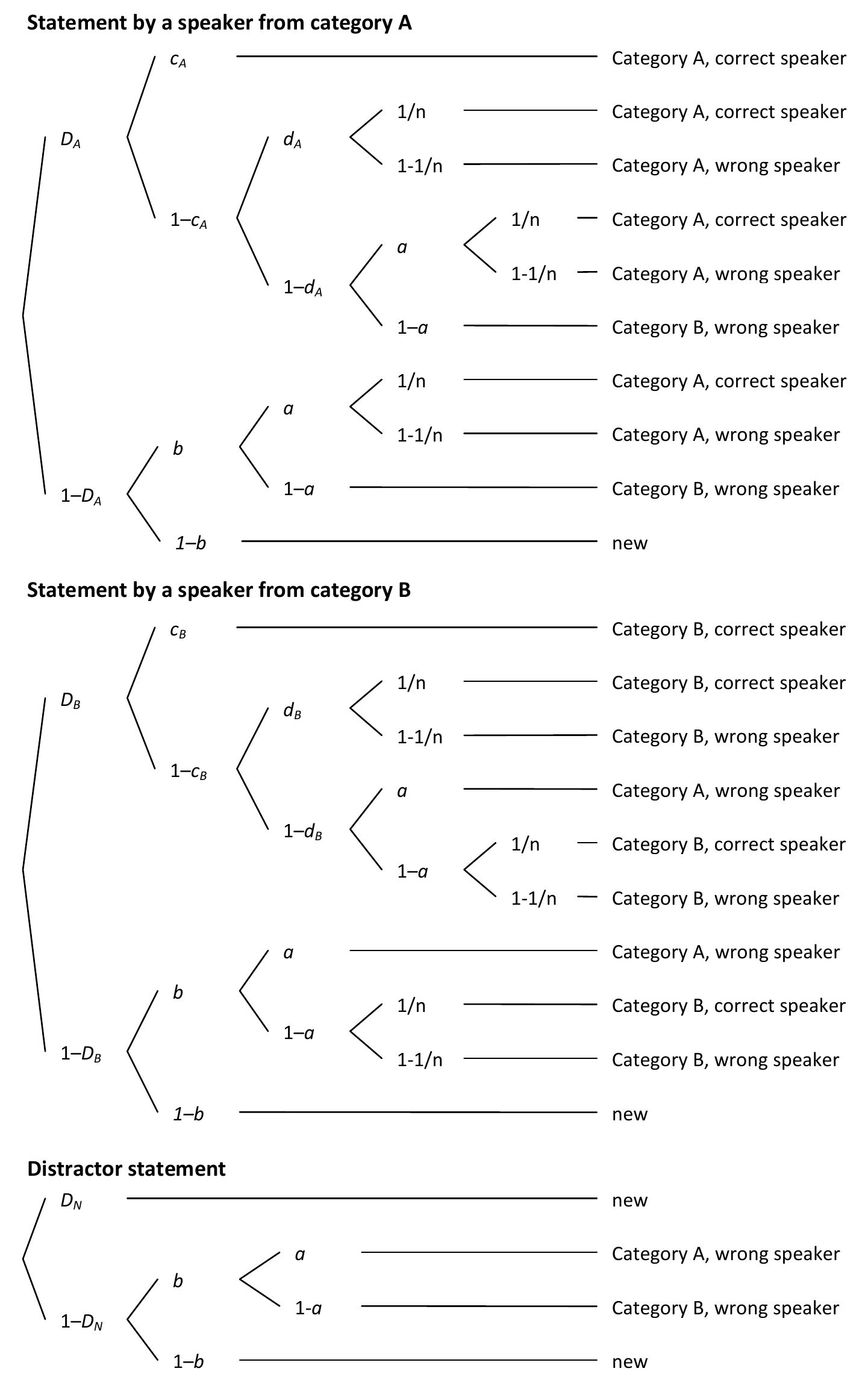}
   \caption{MPT model for the 'Who said what?' paradigm by \protect\citet{Klauer1998}.}
   \label{Fwsw}
\end{figure}

The second example is an MPT model for the “Who said what?” paradigm applied in a social psychological setting \citep[Figure \ref{Fwsw};][]{Klauer1998}. This model is an extension of the two-high-threshold model to measure memory for categories, statements, and persons. As in the studies reported by \citet{Klauer1998}, the nesting model assumes equal probabilities for detecting a statement ($D_A = D_B = D_N$). A nested model additionally constraints the probabilities of discriminating the category of a statement to be equal ($d_A = d_B$). In Table \ref{tabN}, the resulting lower-bound $N'$ is shown as a function of the proportion of distractor items given four persons per category. As for the source-monitoring model, a ratio of 50\% distractors is optimal for the application of FIA ($N' = 2,568$), but still results in a lower-bound $N'$ that is larger than the number of observations of Experiments 2, 3, and 5 (each $n = 1,920$) of \citet{Klauer1998}, for example. Since the fit of the nested model can never compensate for its larger complexity term, the nesting model would always be selected by FIA for these data sets. Again, model selection by FIA would be misleading.

\section{4. Non-nested model comparison}
The third example demonstrates the relevance of our approach for non-nested NML stable models. The models stem from the field of judgment and decision making and describe the behavior of choosing one of two choice options to maximize a given criterion \citep[i.e., decision strategy;][]{Broder2003a}. Recently, \citet{HilbigUR} proposed to classify participants according to strategy usage by means of FIA. Figure \ref{Fjdm} depicts two possible decision strategies, take-the-best (TTB) and a probabilistic weighted-additive rule (WADDprob), each including the error terms $e_1$, $e_2$, and $e_3$ of making a strategy-inconsistent decision. Whereas TTB assumes homogeneous error terms, smaller or equal to chance ($e_1 = e_2 = e_3 \leq 0.5$), WADDprob poses an order restriction on the error probabilities ($e_1 \leq e_3 \leq e_2 \leq 0.5$). Note that TTB and WADDprob are non-nested due to a different definition of the error term $e_2$ ($e_{2, \text{TTB}} = 1 - e_{2,\text{WADDprob}}$). 

Since the number of possible observations increases only according to a cubic law in respect to $N$ (e.g., $\mid \mathcal{X}^N \mid = [(N+1)/3]^3$ for equal proportions of each item type), $C_\text{NML}(N)$ can be computed directly for small to moderate sample sizes without requiring extensive numeric integration techniques. The NML complexity terms are shown in panel A of Figure \ref{Fcfia}. It is evident that the NML complexity curves do not intersect for an equal proportion of item types. Thus, the two non-nested models are NML stable for $N \leq 150$.\footnote{We also checked the NML stability of TTB and WADDprob for other proportions of item types with identical results regarding NML stability .} The respective FIA complexity curves are shown in panel B of Figure \ref{Fcfia}. Comparing the  FIA and NML complexities indicates that FIA reasonably approximates NML for the TTB strategy across all $N$, but strongly underestimates NML for the WADDprob strategy. Thus, in the present example, the observed inversion of FIA complexity terms at $N’ = 80$ can be attributed to a bias in FIA associated with the WADDprob model for small $N$’s. This result also shows that $N > N’$ does not guarantee that FIA approximates NML well, as the bias in FIA for WADDprob is still substantial even for the largest $N$ considered. Nevertheless, ensuring that the number of observations exceeds the lower-bound $N’$ will lead to correct model selections (in terms of NML) and thus still informs the decision to rely on FIA, as severe biases in model selection are avoided and settings are revealed in which direct estimation of NML should be considered. 

In interpreting the resulting lower bound $N’$ of 80 it is important to consider that strategies are usually classified for each participant separately, sometimes based on relevant response frequencies as low as 15 \citep[e.g.,][]{Broder2003b}. Table \ref{tabN} shows that the lower-bound $N’$ cannot be reduced further by changing the proportion of item types. Therefore, for FIA-based model selection to make sense at the individual level, the minimum requirement for \citeauthor*{HilbigUR}’s (in press) candidate models is to obtain at least 27 decisions for each of the three item types per participant.

\section{5. Discussion}
The advantage of considering the functional form in measuring model complexity has often been discussed in the literature \citep[e.g.,][]{Grunwald2000, Myung2006}. Unfortunately, the approximation of NML by means of FIA can be misleading  \citep{Navarro2004}, but no precise condition or model characteristic is known yet to predict or correct this problem. In extreme cases, this problem may result in a reversed rank order of FIA complexity terms for NML stable models, in turn leading to severely biased model comparisons. As a practical solution, we propose to calculate $N'$, the lower-bound sample size for the application of FIA to NML stable model families. If the observed sample size exceeds the lower-bound $N'$, researchers can be confident that the rank order of the $C_\text{FIA}(N)$ complexity terms agrees with the corresponding NML rank order. Otherwise, FIA should not be used, as model selection will be biased in favor of the more complex model in terms of NML. Specifically, in the case of two nested models, the more flexible nesting model will always be selected for $N < N'$, regardless of the data.

We demonstrated the relevance of our approach using three examples in which the FIA criterion results in a misleading model selection even for moderately large sample sizes. However, although $N > N'$ ensures that using FIA does not distort model comparisons, the lower-bound $N'$ cannot be used to determine the required sample size to guarantee a reliable approximation of NML via FIA.

In any case, it is clear that the source of the current difficulty does not lie within the MDL principle itself: The NML criterion does not suffer from this problem at any $N$.  It is also important to keep in mind that implementations of the MDL principle such as FIA provide the advantage of considering the functional complexity of models with the same number of free parameters even if $N$ is small. Such models may include order restrictions and different functional forms that are not captured by information criteria such as AIC and BIC. FIA can readily be applied to compare models with equal numbers of parameters, as the rank order of the complexity terms is always stable because the complexity penalties differ only in the integral. Most importantly in the present context, for NML stable models that differ in the number of parameters, the approach advocated herein provides a safety belt for substantive researchers who want to take advantage of the benefits of FIA while avoiding severe biases in model selection that may occur in case of small samples.


\end{document}